\documentclass[journal=jacsat,manuscript=communication,layout=twocolumn]{achemso}
\usepackage[version=3]{mhchem} 
\usepackage{amsmath,bm,multirow,relsize}

\usepackage{amssymb}
\usepackage{amsthm}
\usepackage{amsmath}

\usepackage{subfigure}

\author{Georgios G. Vogiatzis}
\author{Evangelos Voyiatzis}
\author{Doros N. Theodorou}
\email{doros@central.ntua.gr}
\affiliation[NTUA]{School of Chemical Engineering, National Technical University of Athens, 
Zografou Campus, GR-15780 Athens, Greece}
\title{Monte Carlo simulations of a coarse grained model 
       for an athermal all-polystyrene nanocomposite system}

\makeatletter
\let\thetitle\@title
\let\theauthor\@author
\makeatother

\makeatletter
\renewcommand\section{\@startsection{section}{1}{\z@}%
                                  {-3.5ex \@plus -1ex \@minus -.2ex}%
                                  {2.3ex \@plus.2ex}%
                                  {\normalfont\small\bfseries}}                                  
\makeatother

\makeatletter
\renewcommand\subsection{\@startsection{subsection}{1}{\z@}%
                                  {-3.5ex \@plus -1ex \@minus -.2ex}%
                                  {2.3ex \@plus.2ex}%
                                  {\normalfont\small\small\bfseries}}
\makeatother

\begin{document}

\begin{abstract}
The structure of a polystyrene matrix filled with tightly cross-linked polystyrene nanoparticles, 
forming an athermal nanocomposite system, is investigated by means of a Monte Carlo sampling formalism. 
The polymer chains are represented as random walks and the system is described through a coarse grained Hamiltonian. 
This approach is related to self-consistent-field theory but does not invoke a saddle point approximation and is 
suitable for treating large three-dimensional systems. The local structure of the polymer matrix in the vicinity 
of the nanoparticles is found to be different in many ways from that of the corresponding bulk, 
both at the segment and the chain level. The local polymer density profile near to the particle displays 
a maximum and the bonds develop considerable orientation parallel to the nanoparticle surface. 
The depletion layer thickness is also analyzed. The chains orient with their longest dimension 
parallel to the surface of the particles. Their intrinsic shape, as characterized by spans and principal 
moments of inertia, is found to be a strong function of position relative to the interface. 
The dispersion of many nanoparticles in the polymeric matrix leads to extension of the chains when their size 
is similar to the radius of the dispersed particles.
\end{abstract}

\section{Introduction}
\label{intro}
Nanocomposite polymeric materials consist of nanoparticles dispersed in a polymeric matrix. 
This new family of composite materials displays a variety of properties that attract great scientific and industrial interest. 
The practice of adding nanoscale filler particles to reinforce polymeric materials can be traced back to the early years of the composite industry. 
The design of such conventional composites has focused on maximizing the interaction 
between the polymer matrix and the filler \cite{Gersappe_PRL_2002}. 
This is commonly achieved by shrinking the filler particles in order to increase the surface area available for interaction with the matrix. 
With the appearance of synthetic methods that can produce nanometer sized fillers, resulting in an enormous increase of surface area, polymers reinforced with nanoscale particles should show vastly improved properties. 
Nanomaterials fabricated by dispersing nanoparticles in polymer melts have the potential for performances exceeding those of traditional composites by far \cite{Balazs_Science_2006}.
\par
Even though some property improvements have been achieved in nanocomposites, nanoparticle dispersion is difficult to control, with both thermodynamic and kinetic processes playing significant roles. 
It has been demonstrated that dispersed spherical nanoparticles can yield a range of multifunctional behaviour, including a viscosity decrease, reduction of thermal degradation, increased mechanical damping, enriched electrical and/or magnetic performance, and control of thermomechanical properties 
\cite{Kumar_2006, Youn_JChemPhys_2004, Bockstaller_PRL_2004, Si_Macromolecules_2006, Stratford_Science_2005}. 
The tailor-made properties of these systems are very important to the manufacturing procedure, as they fully overcome many of the existing operational limitations. As a final product, a polymeric matrix enriched with dispersed particles may have better properties than the net polymeric material and can be used in more demanding and novel applications. Therefore, an understanding and quantitative description of the physicochemical properties of these materials is of major importance for their successful production. 
\par
The physical and chemical behavior of this kind of materials depends on the nature of the nanodispersed phase. 
The characteristics of the surface of the particles significantly affect the behavior of the composite polymeric / particle system. 
In some cases, the dispersed nanoparticles in the polymeric matrix may aggregate. 
The presence of particles with grafted chains improves wetting effects in the system and suppresses aggregation, leading the material to dramatically different performance.
\par
Extensive experimental work on all-polystyrene nanocomposites has been recently presented
\cite{Tuteja_PRL_2008, Mackay_Science_2006, Mackay_NatureMaterials_2003}.
Regarding the conformational properties, the dimensions of matrix chains, measured by neutron scattering techniques, 
indicate swelling induced by the dispersed tightly cross-linked nanoparticles. 
This effect occurs only when the radius of gyration of the chain is larger than the nanoparticle radius. 
The all-polymer composite system appears to be a stable blend even for small interparticle distances, 
suggesting that chain shape in the polymer matrix is highly distorted. 
As for the dynamic properties, the blend viscosity was found to demonstrate a non-Einstein-like decrease
with nanoparticle volume fraction. 
It was suggested that this phenomenon scales with the change in free volume introduced by the nanoparticles, while the 
entanglements seem to be totally unaffected.
\par
The effect of incorporating a spherical nanoparticle on the conformational properties of a polymer has been demonstrated 
by means of Monte Carlo (MC) simulations for dilute and semi-dilute solutions  
\cite{Doxastakis_JChemPhys_2004, Doxastakis_JChemPhys_2005}.
These studies have been quite generic, employing a bead-spring and hard-sphere representation for the polymer and the nanoparticle, respectively. 
The most significant changes in structural properties and orientation occurred within the depletion layer. 
The effective interactions between two dispersed nanoparticles in a polymer solution have been quantified 
by calculating the potential of mean force through the expanded ensemble density of states technique \cite{Doxastakis_JChemPhys_2005}. 
An oscillating long-range behaviour has been revealed.
\par
Efforts have also been made to predict the aggregation or the dispersion of polymer nanoparticles in polymer matrix
by thermodynamic modelling. 
Recent advances are thoroughly reviewed in \cite{Hall_CurrentOpinion_2010} and include compressible regular solution 
free energy models \cite{Luzuriaga_JChemPhys_2009,Pomposo_PCCP_2008}, based on modification of theories for binary polymer blends, 
and generalizations of integral equation theories \cite{Jayaraman_Macromolecules_2008, Hall_JChemPhys_2008}, 
such as the microscopic Polymer Reference Interaction Site Model (PRISM). 
The proposed models extend existing theories by incorporating specific nanoparticle-nanoparticle and nanoparticle-polymer contributions. 
The predicted phase diagrams indicate that such systems exhibit a rich variety of behaviours, including upper critical solution temperature-type, lower critical solution temperature-type, and hour-glass shape.
\par
A major challenge in simulating realistic nanocomposite materials is that both the length and the time scales cannot be adequately treated by means of atomistic simulations because of the extensive computational load. 
This is why a variety of mesoscopic techniques have been developed for these particular systems. 
Among them, Self-Consistent Field Theory (SCFT) seems to be a well-founded simulation tool \cite{Fredrickson_EquilibriumTheory}. 
This method adopts a field-theoretic description of the polymeric fluids and makes a saddle-point (mean-field) approximation.
Significant progress has been made in applying the SCFT to nanoparticle-filled polymer matrices in the dilute and semi-dilute 
region \cite{Surve_Langmuir_2006, Surve_Macromolecules_2007, Ganesan_PhysRevE_2008}. 
A lattice-based simulation in the context of the Scheutjens-Fleer approximation has cast some light onto the equilibrium 
dispersion of nanoparticles with grafted polymer chains into dense polymer matrices, where the matrix and the brushes share the 
same chemical structure \cite{Harton_PolymerScience_2008}. 
An attempt to overcome the restrictions posed by the saddle-point approximation was made in \cite{Sides_PRL_2006}. 
The coordinates of all particles in the system were explicitly retained as degrees of freedom. 
It was crucial to update simultaneously the coordinates and the chemical potential field variables.
\par
In the same spirit of relaxing the mean field approximation, Balazs and coworkers have proposed a coupling of SCFT for the inhomogeneous polymers and density functional theory (DFT) for the nanoparticles \cite{Thompson_Macromolecules_2002, Buxton_Macromolecules_2003, Lee_PRL_2003}. 
The extrema of the approximate energy functional created in this way correspond to the mean-field solution. Although the SCFT-DFT approach is quite promising, it is currently  hindered by the difficulty in extending to systems for which reliable density functionals are not available.  
\par 
A novel way of incorporating fluctuation effects in self-consistent mean-field theories has been recently proposed \cite{DaoulasJChemPhys_2006,DaoulasSoftMatter_2006}. 
The ``Single Chain in Mean Field'' simulation technique is particle-based and founded on an ensemble 
of independent chains in fluctuating, external fields. The fields mimic the effect of the instantaneous 
interactions of a molecule with its neighborhood. 
They are frequently updated using the spatially inhomogeneous density distribution of the ensemble. 
The explicit conformations of the molecules evolve by Monte Carlo. 
This technique has been very succesfull in describing the self-assembly and the ordering of block copolymers on patterned substrates encountered in processes such as nanolithography \cite{Detcheverry_PRL_2009,Daoulas_Macromolecules_2008, Harton_PolymerScience_2008, Stoykovich_Macromolecules_2010}. 
\par 
The present work examines the structure of a polymer matrix in the vicinity of polymeric nanoparticles. 
Particular attention is given to capturing the orientational effects and the chain conformation of the 
matrix in a dense, polymeric nano-filled composite. 
The analysis is based on a coarse grained Monte Carlo simulation method which can be 
readily applied to the complex three-dimensional nanocomposite system. 
Although the level of description is quite abstract (i.e. that of the freely-jointed chain for the matrix), the developed model 
tries to predict the behaviour of a nanocomposite with specific chemistry, namely of an all-polystyrene material. 
A main characteristic of the method is that it treats in a different manner the interaction between polymer and between polymer 
and particle: the former is accounted for through a suitable functional of the local densities, 
while the latter is described directly by an explicit interaction potential.

\section{Method}
\subsection{Polymer Coarse Grained Model}
\par 
The aim of the developed coarse grained model is to describe the properties of 
the nanocomposite material at mesoscopic length scales. 
The main parameters are the chain connectivity, the finite compressibility of the 
melt and explicit pairwise potentials for the van der Waals interactions between polymer segment - nanoparticle and 
nanoparticle - nanoparticle.
The system is contained in volume $V$ at temperature $T$. 
The matrix consists of $n$ polymeric chains which are considered to obey the freely jointed chain model. 
Each chain consists of $N$ Kuhn segments.  
The bond length is kept fixed and is equal to the Kuhn length, $b$, of the specific polymer under consideration. 
The $n_{p}$ nanoparticles are represented as impenetrable spheres.
The interaction energy $\mathcal{H}$ includes only non-bonded contributions:
\begin{equation}\label{eq:method_total_interaction}
\mathcal{H} \left( \vec{r} \right)  =
\mathcal{H}_{pp} \left\{\widehat{\rho} \left(\vec{r}\right) \right\} +
\mathcal{H}_{pn} + \mathcal{H}_{nn}
\end{equation}
where $\mathcal{H}_{pp}$ stands for the non-bonded interactions between the polymeric chains, 
$\mathcal{H}_{pn}$ for the interaction between polymeric segments and nanoparticles and
$\mathcal{H}_{nn}$ for the interaction between nanoparticles.

\par 
An important difference in the treatment of polymer-polymer interactions compared to those arising because of 
the presence of the nanoparticle is that the former are taken into account through a suitable functional of the instantaneous 
local density $\widehat{\rho}\left(\vec{r}\right)$, while the latter are accounted for by a pairwise interaction potential. 
The local density is computed directly from the segment positions by a smoothing procedure. 
The non-bonded interactions $\mathcal{H}_{pp}$ are given by:
\begin{equation} \label{eq:method_nonbonded_interactions}
\beta \mathcal{H}_{pp} \left\{\widehat{\rho} \left(\vec{r}\right) \right\} =
\frac{\rho_0}{N} \int_{V} d^3 \vec{r} \left(\frac{\kappa_0 N}{2}
{\left[\widehat{\rho}\left(\vec{r}\right) - 1 \right]}^2  \right)
\end{equation}
where $\beta = \frac{1}{k_BT}$ with $k_B$ being the Boltzmann constant and
$\rho_{0}$ is the average bulk number density of segments. 
Deviations of the local density from the average bulk value are restricted by the quadratic term which stems 
from Helfand's approximation \cite{Helfand_Tagami_JChemPhys_1972a}. 
Thus, the melt has a finite compressibility, which is inversely proportional to $\kappa_0$:
\begin{equation}\label{eq:method_kzero_definition}
\kappa_0 = \frac{1}{k_B \: T \: \kappa_T \rho_0} \;.
\end{equation}

\par 
Although the coordinates of the particles, either polymeric segments or nanoparticles, 
are the degrees of freedom in our model, 
the local densities that appear in \eqref{eq:method_nonbonded_interactions} are not defined by the 
microscopic expression 
$\widehat{\rho} \left(\vec{r}\right) = \sum_{j=1}^{n} \sum_{i=0}^{N} \delta \left(\vec{r} - \vec{r}_{j,i}\right)$. 
Rather, they are obtained by a particle-to-mesh (PM) assignment technique. 
A uniform, rectangular grid with $n_{sites}$ sites is introduced. 
The local densities are defined on each site by applying a smoothing procedure to the bead's position. 
As a consequence, a discretization parameter $\Delta L$ has to be introduced; it is the grid spacing. 
Phenomena and details of the system whose length scales are smaller than the $\Delta L$ parameter cannot be described accurately. 
Thus, this parameter sets the resolution level of our observations. 
It also defines a sort of microscopic cutoff, since it determines the range of interaction between neighboring beads. 
It follows that $\Delta L$ must not be less than the average distance between two beads. 
The value at which $\Delta L$ should be set depends strongly on the system of interest and the preferred smoothing procedure.

\par
\begin{figure}
\centering
\includegraphics[width=\columnwidth]{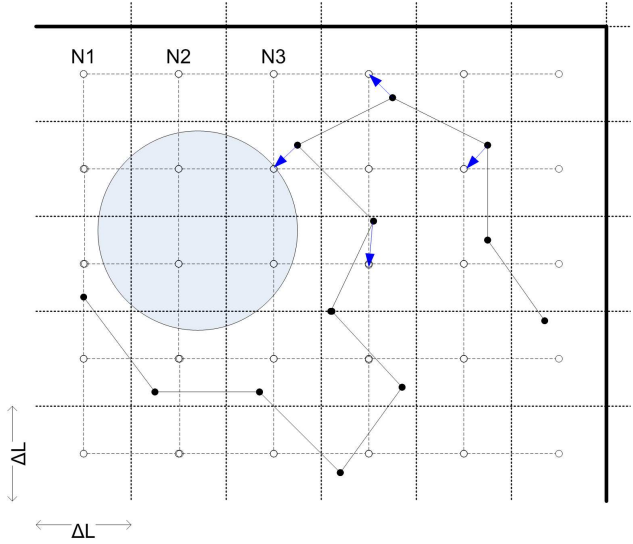}
\caption{Schematic of the PM technique used to estimate the local densities.}
\label{fig:method_assignment}
\end{figure}

 
The assignment procedure of a polymeric bead to the sites of the rectangular grid is shown in Figure \ref{fig:method_assignment}.
The $i$-th segment of the $j$-th chain at position $\vec{r}_{j,i}$ shown as a black dot, 
yields a contribution to a neighboring $m$-th site, shown as white dot, 
$\Pi \left[\vec{r}_{j,i}, \vec{c}_m \right]$, 
where $\vec{c}_m$ is the position of the centre of the $m$th site and $\Pi$ the assignment function.
The resulting normalized density at site $m$ is calculated as the sum of the contributions of all segments in the system,
giving the instantaneous number of segments assigned to the site:
\begin{equation} \label{eq:method_cell_density_function}
\widehat{\rho}_m = \frac{1}{\rho^{nor}_{m}} 
\sum_{j=1}^{n}
\left\{
\sum_{i=0}^{N} \Pi \left[\vec{r}_{j,i}, \vec{c}_m \right] 	\right\} 
\left[1 -\frac{1}{2} \left(\delta_{j,0} + \delta_{j,N} \right) \right] 
\end{equation}
with $\delta$ the Kronecker delta.

\par 
The normalization constant $\rho^{nor}_{m}$ represents the segment number density that the site $m$ would have, 
provided that the melt was completely homogeneous without any density fluctuation, $\rho^{nor}_{m} = nN / n_{sites}$. 
The assignment function $\Pi$ could be chosen arbitrarily. 
In the present study, a zero-order (PM0) interpolation scheme is employed. 
This choice corresponds to the case where the bead is assigned entirely to its closest site and does not contribute 
to any other site. It follows that two beads interact only when they are assigned to the same site. 
It is of great importance to use a dense grid with many sites, so as to minimize the discretization 
effect on the density. 

\par 
Apart from the polymer-nanoparticle pairwise potential, 
the mere existence of the nanoparticle, which is an impenetrable spherical object, defines a second, 
implicit way of interaction with the matrix. 
The available volume that the polymer can occupy in cells, that are in the vicinity of the nanoparticle, 
is less than the volume which can be occupied in cells far away from it. Since the volume of the cells touching the 
nanoparticle is truncated, the normalizing density $\rho^{nor}_m$ of an arbitrary cell should be given by:  
\begin{equation} \label{eq:method_cell_normalizing_constant}
\rho^{nor}_m = \rho_0 \int_{V^{acc}_m} d\vec{r} \; \Pi \left[\vec{r}_{j,i}, \vec{c}_m \right]
\end{equation}
where $V^{acc}_m$ is the volume of the cell space accessible to the polymeric segments. 
For the case of the PM0 scheme, $\rho^{nor}_m$ is simply proportional to the accessible (non nanoparticle-occupied)
volume associated
with site $m$, $\rho^{nor}_m = \rho_0 V^{acc}_m$. 
The calculation of the accessible free volume of each cell of the grid is based on a suitable adaptation of 
an analytical algorithm \cite{DoddTheodorou_MolPhys_1991}. 
The original algorithm determines in an analytical fashion the volume of a sphere delimited by a set of arbitrary directed planes. 
It has been adapted to treat the case of the volume of a sphere intersecting with a 
cubic box and takes into account the periodic boundary conditions. 
The same strategy can be applied to the case of multiple particles intersecting one box, since the former cannot 
intersect each other and the available volume can be readily estimated by subtracting the  
volume occupied by each particle.
\par 
An alternative to using a grid in order to estimate the local densities is to resort to cloud-density 
methods \cite{Laradji_PhysRevE_1994}. 
Such techniques call for a process that identifies the pair of beads which are close enough to interact. 
A discretization parameter closely related to $\Delta L$ employed in the PM case, the Gaussian width, 
has to be defined. There is also a significant increase of computational load. 
Despite the demanding nature of the cloud methods, the accuracy of their predictions has been shown to be the same as that of 
grid-mediated ones and both methods give essentially the same results \cite{Laradji_PhysRevE_1994, Detcheverry_PRL_2009}. 
 
\subsection{Nanoparticle-polymer and nanoparticle-nanoparticle interaction}
The effect of dispersing bare nanoparticles in the polymer matrix is explicitly accounted. 
Pairwise potentials describing the effective interaction between polymeric segment - nanoparticle, $U_{pn}^{eff}$, and
nanoparticle - nanoparticle, $U_{nn}^{eff}$ are introduced. 
The overall explicit energetic effect is taken into account in the Hamiltonian by summing up these 
interactions for all possible pairs of nanoparticles and polymeric segments:
\begin{equation} \label{eq:eff_nano_polymer}
\mathcal{H}_{pn} = \sum_{i_n=1}^{n_p} \; \sum_{i_p=1}^{nN} U_{np}^{eff} \left(\vec{r}_{i_n} - \vec{r}_{i_p} \right)
\end{equation} 
and
\begin{equation} \label{eq:eff_nano_nano}
\mathcal{H}_{nn} = \sum_{i_n=1}^{n_p-1} \; \sum_{j_n=i_n+1}^{n_p} U_{nn}^{eff} 
\left(\vec{r}_{i_n} - \vec{r}_{j_n} \right) \;.
\end{equation} 

\par
\begin{figure}
\centering
\includegraphics[width = \columnwidth]{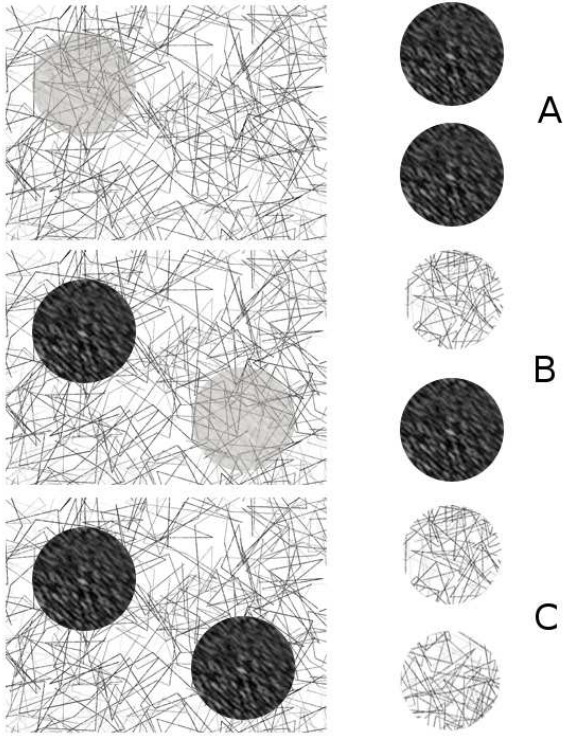}
\caption{Insertion of two nanoparticles in the matrix, in three discrete steps. 
In step A a spherical volume of polymeric matrix is removed from the melt.
In step B the first spherical particle is placed in the place of the removed volume of polymeric matrix. Finally in step C
another volume of polymeric matrix is replaced by the second nanoparticle.}
\label{fig:flocculation_steps}
\end{figure}

 
The reference energy level of the employed Hamiltonian \eqref{eq:method_nonbonded_interactions} for the homopolymer melt 
is that of a melt with uniform density profile. In this case, the effective interaction energy 
between two polymeric beads is taken as zero. 
The effective potentials introduced in \eqref{eq:eff_nano_polymer} and \eqref{eq:eff_nano_nano} should be such that, if 
the volume of the nanoparticles was occupied by bulk homopolymer, then the energy of the system would be zero. 
The insertion of a nanoparticle in a uniformly-distributed melt can be thermodynamically accomplished in two steps: 
a spherical volume of polymer, whose net interaction with the rest of the polymer matrix is $U_{pp}^{equiv-p}$, 
is removed from the melt and a nanoparticle with equal volume is placed in its position, introducing a new net 
interaction $U_{pn}$ with the remaining bulk polymer (Figure \ref{fig:flocculation_steps} from step A to step B). 
Thus, the net interactions between polymer-polymer should be subtracted from the net interactions 
between polymer-nanoparticle \cite{Hiemenz}:
\begin{equation} \label{eq:method_nano_poly_interaction}
U_{pn}^{eff} = U_{pn} - U_{pp}^{equiv-n} \;\;.
\end{equation}

\par 
The treatment of the effective interactions between two nanoparticles is more sophisticated. 
Their insertion in the melt, following the above thermodynamic way, requires two more additional steps 
with respect to the previous case. 
A second  spherical volume of polymer is removed from the melt in the presence of one nanoparticle,
whose net interaction with the rest of the system is $U_{pn}^{equiv-n}$,  
and the second particle is inserted in the created volume, 
introducing new net interaction $U_{nn}$ with the composite system (Figure \ref{fig:flocculation_steps} from step B to step C). 
\begin{equation} \label{eq:method_nano_nano_interaction}
U_{nn}^{eff} = U_{nn} -2 U_{pn}^{equiv-n} + U_{pp}^{equiv-n}
\end{equation}

\par
\begin{figure}
\centering
\includegraphics[width=\columnwidth]{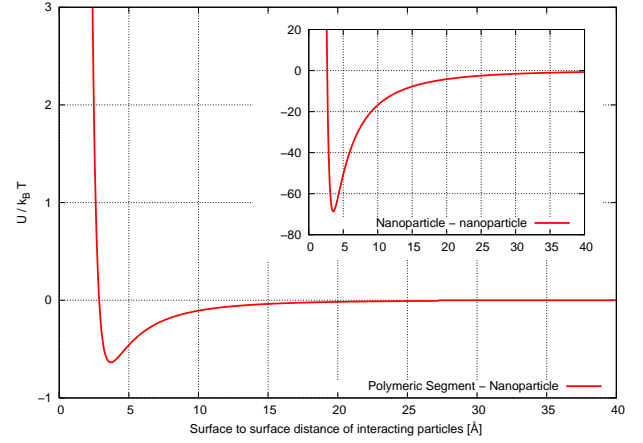}
\caption {Pairwise interaction potentials used for polymeric segment and nanoparticles with the energy axis 
scaled by $k_B T$. The density of the nanoparticle is $1.15 \frac{\text{g}}{\text{cm}^3}$ and
its radius is $R_n = 36$ \AA. 
The radius of the polymeric segment interaction site $R_b$ is $6.7$ \AA. 
The centre-to-centre distance is shifted by $R_n+R_b$ in the case
of the polymeric segment-nanoparticle interaction (main figure) 
and by $2R_n$ in the case of nanoparticle - nanoparticle interaction (inset).}
\label{fig:potential_plot}
\end{figure}

 
The coarse grained potentials describing the net interactions involved in 
\eqref{eq:method_nano_poly_interaction} and \eqref{eq:method_nano_nano_interaction} are derived on the 
grounds of Hamaker theory \cite{Hamaker_original}. 
A major advantage is that the existence of accurate atomistic potentials describing these interactions is exploited. 
The need to introduce empirically adjustable parameters is kept to a minimum. 
The integrated potentials are computationally effective. 
Each particle is treated as a collection of atoms which interact only with atoms of different particles:
\begin{equation}\label{eq:method_general_potential_integration}
U\left(r_{12}\right) = \int_{sphere1} \int_{sphere2} \rho_1 \left(\vec{r}_1 \right)
\rho_2 \left(\vec{r}_2 \right) U_{LJ}\left(r_{12}\right) dV_1 \: dV_2 
\end{equation}
where $\rho_1$ and $\rho_2$ stand for the density of interaction centres in each particles, $r_{12}$ is the distance between the
centres of the particles, and $dV_{1},dV_{2}$ the elementary volumes of each particle.
The discrete distribution of the interacting atoms in each particle is assumed to be continuous and uniform. 
The shape of the particles is spherical. 
An analytical closed-form solution can be rigorously derived,  
 provided that the atomistic potential is  Lennard-Jones 12-6 \cite{Hamaker_original, Everaers_2003}:
\begin{eqnarray}
U \left(r_{12} \right) & = & U_{A} (r_{12}) + U_R (r_{12}) \\
U_A (r_{12}) & = & -\frac{A_{12}}{6} \Bigg[\frac{2\alpha_1 \alpha_2}{{r_{12}}^2 - \left(\alpha_1 + \alpha_2 \right)^2} 
+ \frac{2\alpha_1 \;\alpha_2}{{r_{12}}^2 - \left(\alpha_1 - \alpha_2 \right)^2} \nonumber \\ 
& + & \ln \left( \frac{{r_{12}}^2 - \left(\alpha_1 + \alpha_2 \right)^2}
{{r_{12}}^2 - \left(\alpha_1 - \alpha_2 \right)^2} \right) \Bigg] \\
 U_R(r_{12}) & = &  \frac{A_{12} \sigma_{LJ}^6}{37800 r_{12}}  \nonumber \\
& \Bigg[ & \frac{{r_{12}}^2 - 7 r_{12} \left(\alpha_1 + \alpha_2 \right) 
+ 6 \left({\alpha_1}^2 + 7\alpha_1\alpha_2 + {\alpha_2}^2\right)}
{\left( r_{12} - \alpha_1 -\alpha_2 \right)^7} \nonumber \\
& + &\frac{{r_{12}}^2 + 7 r_{12} \left(\alpha_1 + \alpha_2 \right) 
+ 6 \left({\alpha_1}^2 + 7\alpha_1\alpha_2 + {\alpha_2}^2\right)}
{\left( r_{12} + \alpha_1 +\alpha_2 \right)^7} \nonumber \\
& - &\frac{{r_{12}}^2 + 7 r_{12} \left(\alpha_1 - \alpha_2 \right) 
+ 6 \left({\alpha_1}^2 - 7\alpha_1\alpha_2 + {\alpha_2}^2\right)}
{\left( r_{12} + \alpha_1  -\alpha_2 \right)^7} \nonumber \\
& - &\frac{{r_{12}}^2 - 7 r_{12} \left(\alpha_1 + \alpha_2 \right) 
+ 6 \left({\alpha_1}^2 - 7\alpha_1\alpha_2 + {\alpha_2}^2\right)}
{\left( r_{12} - \alpha_1 +\alpha_2 \right)^7} \Bigg] \nonumber \\
& &
\end{eqnarray}
where $A_{12} = 4 \pi^2 \varepsilon_{LJ} \rho_1 \rho_2 \sigma_{LJ}^6$ is the Hamaker constant, 
$\alpha_1, \alpha_2$ the radii of the spherical particles, $r_{12}$ the distance between their centres
and $\varepsilon_{LJ}, \sigma_{LJ}$ the  Lennard-Jones energy and distance cross interaction 
parameters for the pair of interaction sites.
In the following, the radius of the particle representing a polymeric segment, 
$R_b$, is chosen such that $(4/3) \pi R^3_b$ equals the volume per Kuhn segment in the bulk polymer, 
as calculated from the mass corresponding to a Kuhn segment and the experimental mass density (approximately $R_b = 6.7$ \AA). 
The closest distance of approach is equal or greater than the radius that is used for the estimation 
of the Hamaker interaction between the nanoparticle and the Kuhn segment.
The atomistic parameters for CH(aromatic), CH$_2$(aliphatic), CH(aliphatic), and CH$_3$(aliphatic) sites 
interacting with each other are taken from \cite{Spyriouni_Macromolecules_2007}. 
The cross-interaction parameters are estimated by a geometric mean combining rule \cite{GoodHope_1971}.
Both nanoparticle - polymeric segment and nanoparticle - nanoparticle potential are depicted in Figure
\ref{fig:potential_plot}.

\subsection{Details of the Monte Carlo Simulations} 
\begin{figure}
\centering
\includegraphics[width=\columnwidth]{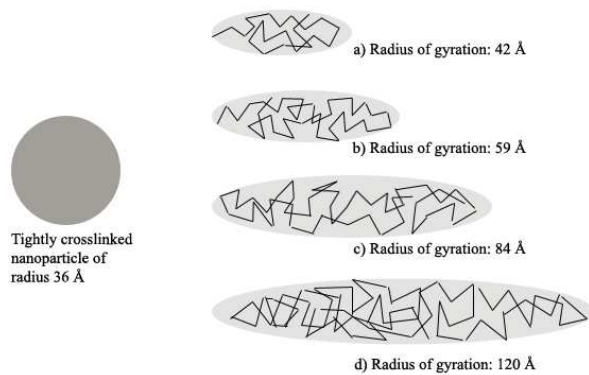}
\caption{Schematics of the relative size of nanoparticle and the four chain lengths under consideration.
For each length the $R_{g0}$ in pure melt is used for the comparison. The chain lengths are $32$ (a), $64$ (b),
$128$ (c) and $256$ (d) Kuhn segments per chain respectively.}
\label{fig:nanoparticle_vs_chains}
\end{figure}

All simulations were carried out in the canonical statistical ensemble (NVT). 
The temperature of the system was $400$ K. 
The box was cubic and its edge length was $600$ \AA . 
It should be noted that, for all chain lengths, the edge length is greater than $4R_g$ with $R_g$ being 
the radius of gyration of a chain, so as to avoid finite 
size effects \cite{Doxastakis_JChemPhys_2004}. 
The systems studied were strictly monodisperse, while the degree of polymerization of the 
chains $N$ was varied, with $N = 32, 64, 128$ and $256$ beads (Figure \ref{fig:nanoparticle_vs_chains}.
The corresponding molecular weights are $23$, $47$, $93$ and $187$ kg/mol, respectively).  
The number of chains in the systems is varied according to chain length so that the mean density of the 
polymer in the accessible volume is $0.97$ $\text{g}/\text{cm}^3$.  
The Kuhn segment length $b$ is $18.3$ \AA, which is assigned to seven styrene monomers. 
The radius of the nanoparticle is $36$ \AA. 
The density of the crosslinked polystyrene nanoparticle is $1.15 \text{g}/\text{cm}^3$. 
The characteristics of the systems studied, i.e. the degree of polymerization of the chain, 
the radius and the mean density of the crosslinked polystyrene nanoparticle, 
are in close connection with systems that have been studied experimentally by MacKay and co-workers
\cite{Tuteja_PRL_2008, Mackay_Science_2006}.

\par 
In our initial calculations, a single nanoparticle is placed at the centre of the simulation box. 
The polymer chains are built randomly \cite{Marsaglia_AnnMathStat_1972} with the constraint that none of 
the polymeric beads overlaps with the nanoparticle. 
When more than one particles had to be contained in a configuration, then the nanoparticles 
were placed first at randomly selected positions, so that they did not overlap, and then the polymer was built around them. 
The initial configurations created by this method have large local density fluctuations. 
These fluctuations increase with increasing chain length. 
A zero temperature Monte Carlo optimization procedure took place in order to reduce the density fluctuations. 
During this stage, all moves leading to more uniform density profile, thus decreasing the density fluctuations, are accepted. 
In the opposite case, they are rejected \cite{AuhlEveraers_JChemPhys_2003}. 
Five intermolecular moves were employed which treat the entire chains as rigid bodies: 
\begin{itemize}
\item
\textit{Translation} of individual chains in a random direction
\item
\textit{Rotation} of individual chains by random angles around random axes through their centres of mass
\item
\textit{Reflection} of individual chains at random mirror planes going through the centre of mass
\item
\textit{Inversion} of individual chains at their centres of mass
\item
\textit{Exchange} of two chains preserving the centre of mass positions
\end{itemize}
When the Monte Carlo proceeds, the internal shape of the chains is restructured by four intramolecular moves: 
the flip, the end rotation, the reptation and the pivot moves. 
The exact mix of moves and acceptance rates of the moves are given in Table \ref{tab:mc_moves_mixture}
for pure melt polystyrene system. 
The nanoparticles are allowed to translate in random directions every $500$ MC steps. 

\begin{table}
\centering
\caption{Mixture of MC moves used for equilibration of pure melt of chains with $32$ polymeric segments per chain}
\label{tab:mc_moves_mixture}
\begin{tabular}{| l | c | r |}
\hline
Move & \% Attempted &  \% Accepted \\
\hline
Rigid translation & 8 \%  &   62.3 \%    \\
Rigid rotation    & 8 \%  &   42.7 \%    \\
Rigid reflection  & 8 \%  &   39.9 \%    \\
Rigid inversion   & 8 \%  &   37.5 \%    \\
Rigid exchange    & 8 \%  &   35.5 \%    \\
Flip              & 15 \% &   77.5 \%    \\
End rotation      & 15 \% &   70.4 \%    \\
Reptation         & 15 \% &   61.8 \%    \\
Pivot             & 15 \% &   53.9 \%    \\
\hline
\end{tabular}
\end{table}

\par
\begin{figure}
\centering
\includegraphics[width=\columnwidth]{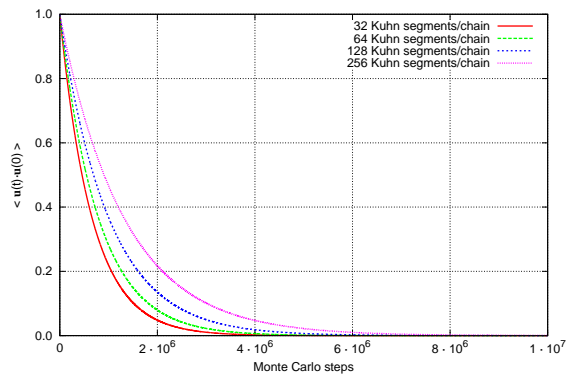}
\caption{Decay of the chain end-to-end vector orientational autocorrelation function $\left \langle \vec{u}(t) \cdot \vec{u}(0) \right \rangle$ for four systems of different chain length at bulk polymer density $0.97 \frac{\text{g}}{\text{cm}^3}$.
The radius of the nanoparticle is $36$ \AA.}
\label{fig:autocorrelation_function}
\end{figure}

 
To demonstrate the efficiency of the MC method in equilibrating the systems under study, 
the orientational autocorrelation function $\left \langle \vec{u} (t) \cdot \vec{u}(0) \right \rangle$ of a unit vector directed
along the chain end-to-end vector was evaluated as a function of number of simulation steps.
Figure \ref{fig:autocorrelation_function} shows the results for the four chain lengths considered here.
The shorter chain systems exhibit faster equilibration rate due to the increased acceptance of rigid 
moves. Rigid moves of smaller chains lead to drastic variation of density fluctuations without affecting dramatically the overall
energy of the grid. As larger molecules are considered, moves of this kind become less drastic because
they tend to create larger density fluctuations (since a large number of segments change positions at the same time). 
On the other hand, intramolecular moves and especially the pivot move become very efficient for large chains,
because of their ability to rearrange a specific part of the chain. The total length of simulations varied between $250$ and
$800$ millions iterations, one order of magnitude above the time needed for the end-to-end vector orientational autocorrelation
function to drop to zero.

\section{Results}

\subsection{Local polymer structure in the presence of a nanoparticle}
\par
\begin{figure}
\centering
\includegraphics[width=\columnwidth]{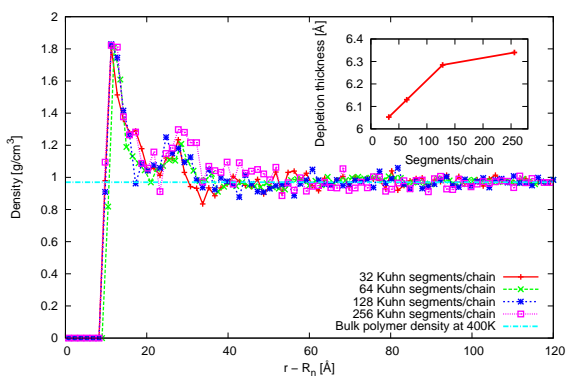}
\caption{Radial mass density distribution from the surface of the nanoparticle accumulated in $1.5$-\AA-thick bins,
for $N = 32$ (continuous line), $N = 64$ (long-dashed line), $N = 128$ (short-dashed line) and $N = 256$ (dotted line).
The radius and the mass of the nanoparticle are $36 \: \text{\AA}$ and $135 \: \text{kDa}$, respectively. 
The straight dot-dashed line represents the bulk polymer density at $400 \: \text{K}$.
In the inset, the variation of the depletion layer thickness $\delta$ with chain length is shown.}
\label{fig:radial_mass_density}
\end{figure}

The local structure in the vicinity of the nanoparticle at the Kuhn segment level 
is first studied through the radial mass density distribution from the surface of the
nanoparticle, which is displayed in Figure \ref{fig:radial_mass_density}. 
The beads of the polymer have been classified in $1.5$-\AA-thick spherical shells whose origin is nanoparticle's centre. 
The distance of the closest segment from the nanoparticle surface is approximately equal to $7$ \AA, 
which corresponds to the radius of the segment's Hamaker interaction site.
Far away from the nanoparticle, the density assumes its bulk experimental value for all chain lengths. 
In the vicinity of the nanoparticle there is a strong maximum that reaches approximately 
$1.8 \frac{\text{g}}{\text{cm}^3}$, located at the distance where appears the minimum of the Hamaker potential . 
This value is observed for all chain lengths. 
The enhancement of polymeric segment density is very similar between the four chain lengths under consideration.
Although being close to the nanoparticle leads to a decrease of the chain entropy 
(allowed conformations are fewer than in the bulk), attractive energetic 
contributions due to the higher density of the nanoparticles overcome these entropic contributions and lead 
to an increase in local density. 
Monomer packing effects are important, since monomers form layers around the nanoparticle. 
A second peak is also observed in the radial density distribution of segments 
and the position in each case does not depend on the number of Kuhn segments of the chain. 
This second peak is located roughly one segment diameter away from the first peak.  
Similar studies in the semi-dilute region \cite{Doxastakis_JChemPhys_2004} have also shown that, 
as the polymer density was increased, the initially 
chain length-dependent peaks become independent. On the basis of the perturbations in the total density profile from its bulk value, 
the thickness of the interface could be estimated as less than $40$ \AA. 
The typical length scale characterizing the density distribution in the vicinity of 
the nanoparticle is the depletion layer thickness $\delta$, which is shown in 
the inset of Figure \ref{fig:radial_mass_density}. 
The depletion layer can be defined as the zone next to the particle where there are no polymer monomers 
\cite{Aarts_JPhysCondMatter_2002}.
It is an equimolar dividing surface. 
The number of segments depleted is estimated by an appropriate mass balance and leads to an expression of the form:
\begin{equation}
(R_n + \delta)^3 - R_n^3 = 3 \int_{R_n}^{\infty} r^2 \cdot \left( 1 - g(r)\right) dr 
\end{equation}
where $R_n$ is the radius of the nanoparticle and $g(r) = \rho\left(r\right) / \rho_0$ is the 
pair distribution function between the nanoparticle
and the polymeric segments around it.
As shown in the inset to Figure \ref{fig:radial_mass_density}, the thickness of the depletion layer 
increases slightly with increasing the number of Kuhn segments in a chain. 
The value is close to the radius calculated for the segment's spherical interaction site. 

\par
\begin{figure}
\centering
\includegraphics[width=\columnwidth]{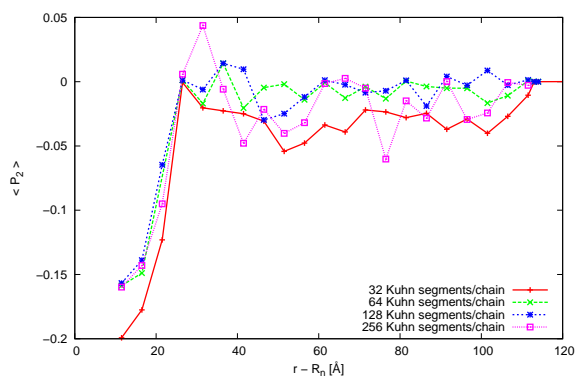}
\caption{Average second Legendre polynomial $P_2$ plotted against the distance of segment $i$ from the surface of the
nanoparticle in $5$-\AA-thick bins. The $P_2$ polynomial is calculated between the vector connecting the centre of
the nanoparticle and a Kuhn segment $i$ and the bond vector of Kuhn segments $i$ and $i+1$.
The radius of the nanoparticle is $36$ \AA. A positive (negative) order parameter indicates an orientational tendency that is
perpendicular (parallel) to the nanoparticle's surface.}
\label{fig:segment_legendre}
\end{figure}

The local orientation of chain segments induced by the nanoparticle can be quantified by the second order 
Legendre polynomial $P_2$. 
An angle $\theta$ may be defined by the bond vector of two consecutive Kuhn segments $i$ and $i + 1$ and 
the vector connecting the centre of the nanoparticle with the Kuhn segment $i$. 
This angle is used to define the $P_2$ function: $P_2 = \frac{1}{2} \left( 3 \langle \cos^2\theta \rangle - 1 \right)$. 
$P_2$ would assume values of $-0.5$, $0.0$, and $1.0$ for bonds characterized by perfectly parallel, random, and 
perpendicular orientation relative to the surface of the particle, respectively. 
Figure \ref{fig:segment_legendre} displays the radial distribution of $P_2$ from the centre of a 
nanoparticle, accumulated in $5$-\AA-thick bins. 
A given bond is assigned to the bin in which its $i$ Kuhn segment resides. 
The bonds lying in the vicinity of the particle are structured and tend to be oriented tangentially 
to the interface. 
Chain length effects on the orientational distribution of bonds are weak.  
The tangential orientation tends to be stronger for the shortest chains examined.
The shorter the chain, the more intense the orientational effects are. 
Kuhn segment orientation effects are weak beyond a distance of two Kuhn segment lengths from the particle surface.
Similar behavior has been observed in molecular dynamics simulations of silica particles dispersed in a
polyethylene-like matrix \cite{Brown_Macromolecules_2008}.

\par
\begin{figure}
\centering
\includegraphics[width=\columnwidth]{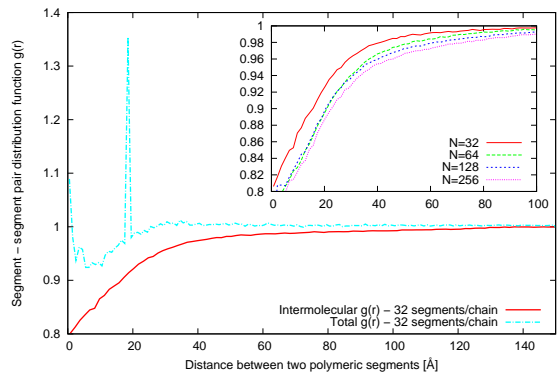}
\caption{Polymeric segments, total and intermolecular pair distribution functions, accumulated in $1$-\AA-thick bins. 
One chain length is depicted, $N=32$ (main figure).
The intermolecular pair distribution function, in the presence of a nanoparticle of radius $36$ \AA, is included in the inset.}
\label{fig:inter_gofr}
\end{figure}

One of the most important aspects of structure in polymer melts is the so-called ``correlation hole'' effect . 
It arises in dense polymeric fluids because of the melt's incompressibility.
Figure \ref{fig:inter_gofr} shows the variation of the total and the intermolecular pair distribution functions
between polymeric segments.
The correlation hole effect is evident in the intermolecular distribution functions. 
On very short length scales, the total distribution function diverges as $1/r$ 
because of the intramolecular contribution dictated by the freely jointed model. 
This is an artificial effect, caused by the fact that chain self-intersection is not prevented in the freely 
jointed chains used in our calculations. 
In the regime of small distances, segments of other molecules are expelled from the volume of a reference 
chain and this gives rise to a correlation hole in the intermolecular pair distribution function.
The spike in the total pair distribution function is an intramolecular feature, caused by pairs of segments 
connected directly by a Kuhn segment. If an atomistic description were used, this spike would be replaced by a 
series of intramolecular peaks reflecting the bonded geometry and conformational preferences of polystyrene molecules.
On the other hand, for large length scales, the intermolecular distribution function approaches unity, 
as the intramolecular distribution function approaches zero.
As shown in the inset of Figure \ref{fig:inter_gofr}, longer chains exhibit ``correlation hole'' effect for slightly 
longer distances.
  
\subsection{Chain structure in the presence of a nanoparticle}
\begin{figure}
\centering
\includegraphics[width=\columnwidth]{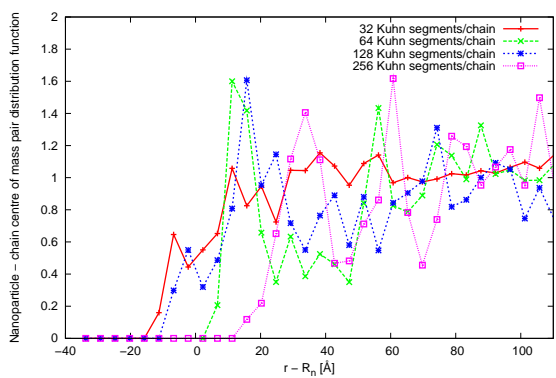}
\caption{Chain centre of mass distribution, accumulated in $4.5$-\AA-thick bins,
for $N = 32$ (continuous line), $N = 64$ (long-dashed line), $N = 128$ (short-dashed line) and $N = 256$ (dotted line).}
\label{fig:cms_around_particle}
\end{figure}

Complementary to the study of local structure at the segment level, long range structural 
features may be revealed by probing properties at the level of entire chains. 
The distribution of the centre-of-mass (COM) of the entire chains around the nanoparticle, which is shown 
in Figure \ref{fig:cms_around_particle}, is such an example. 
The COMs of the chains are classified in $4.5$-\AA-thick bins. 
It should be noted that the results at the level of the entire chains are more prone to 
statistical noise, because the number of binned chains is considerable smaller than the number of polymeric beads. 
There are polymeric chains whose COM position lies inside the nanoparticle. These chains engulf the nanoparticles. 
The shorter the polymeric chains, the more intense the penetration of the nanoparticles into the chains. 

\par
\begin{figure}
\centering
\subfigure[]{
	\label{fig:spans}
	\includegraphics[width=\columnwidth]{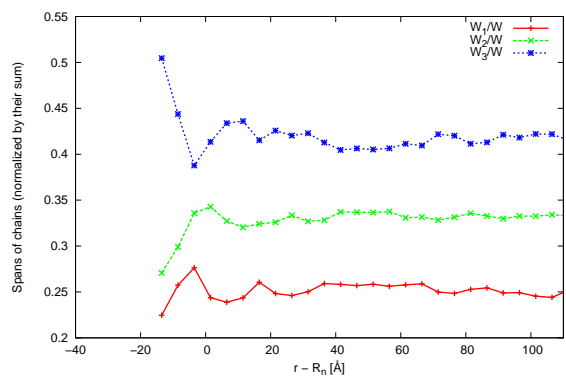}
}
\subfigure[]{
	\label{fig:rg_eigenvalues} 
	\includegraphics[width=\columnwidth]{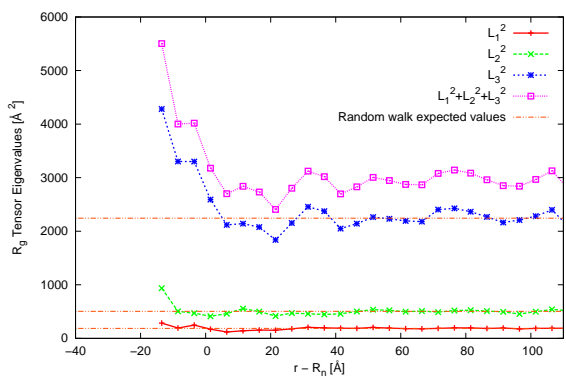}
}
\caption{Overall shape of chains as a function of centre of mass position.
(a,top) Spans of chains, normalized by their sum $W = W_1 + W_2 + W_3$. (b, bottom) Eigenvalues of the 
chain radius of gyration tensor ${L_1}^2, {L_2}^2, {L_3}^2$. These plots reveal that the overall shape of chain 
segment cloud is dependent on position relative to the nanoparticle.
The systems consists of chains with $32$ Kuhn segments per chain and one nanoparticle of radius $36$ \AA.}
\label{fig:spans_eigenvalues}
\end{figure}

The overall shape of the chains is explored by determining their spans and the eigenvalues of their radius of gyration
tensors. Spans are defined according to Rubin and Mazur \cite{RubinMazur_1977}: they are the dimensions of the smallest
orthorhombic box that can completely enclose the chain segment cloud. Our analysis, presented for spans
in Figure \ref{fig:spans}, leads to the conclusion that the overall shape of the 
segment cloud is strongly affected by the presence
of the spherical nanoparticle. Near the surface of the nanoparticle,
chains tend to expand along their main axis (largest span, $W_3$, increases) and shrink along the 
smaller ones (spans $W_2$ and $W_1$ decrease).
The same tendency is revealed through an examination of the eigenvalues of the chain radius of gyration tensor.
In the polymer melt, polymers have the shape of flattened ellipsoids \cite{TheodorouSuter_1985}. For an ellipsoid 
formed by a random walk model, the ratio of the three eigenvalues is reported to be $12:2.7:1$ \cite{PicuOzmusul_2003}.
In our simulations, even the smallest chains are in excellent agreement with this prediction, 
as far as the small and the medium ellipsoidal semiaxes ($L_1$ and $L_2$) are concerned. 
At shorter distances from the nanoparticle centre than the mean size of the chain, 
an expansion of the chains across their principal semiaxis ($L_3$) is
found. This leads to an increase of radius of gyration $R_g^2 = {L_1}^2 + {L_2}^2 + {L_3}^2$ near the nanoparticle. The
deformation of the molecules is smaller for longer chains (whose dimensions exceed by far the radius of the nanoparticle).
Far from the surface of the nanoparticle, the chain dimensions, estimated either as spans 
or as radius of gyration tensor eigenvalues, reach their 
bulk average values, since the molecules are not affected by the presence of the nanoparticle.

\par
\begin{figure}
\centering
\subfigure[]{
	\label{fig:spans_orientation}
	\includegraphics[width=\columnwidth]{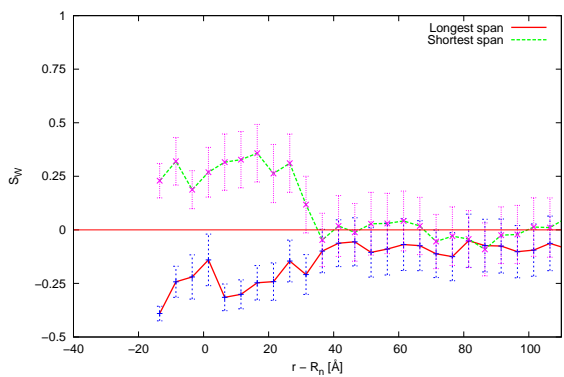}
}
\subfigure[]{
	\label{fig:eigenvalues_orientation}
	\includegraphics[width=\columnwidth]{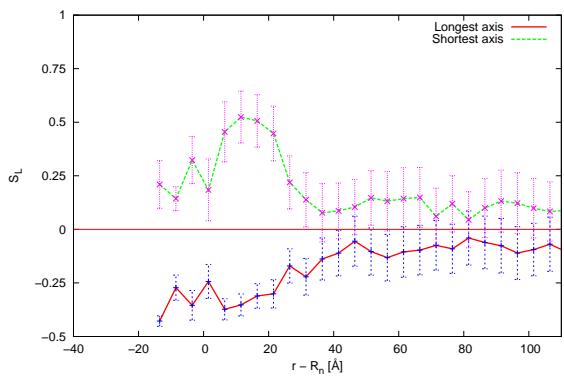}
}
\caption{(a,top) Order parameters for the longest and shortest spans of chains as functions of centre
of mass position. (b,bottom) Order parameters for the eigenvectors corresponding to the largest and the smallest
eigenvalues of the chain radius of gyration tensor as functions on the centre of mass position.
The systems consists of chains with $32$ Kuhn segments per chain and one nanoparticle of radius $36$ \AA.}
\label{fig:order_spans_eigenvalues}
\end{figure}

The orientation of chain segment clouds with respect to the surface of the nanoparticle is explored
by computing order parameters for chain spans and radius of gyration tensor eigenvectors
\cite{TheodorouMansfield_1990, TheodorouMansfield_1991}.
The definition of the order parameter for spans is analogous to that used here for the segments
$S_W = (1/2) \left[3 \langle \cos^2(\theta) \rangle -1 \right]$. The angle $\theta$ is 
formed between the larger span $W_3$ and the vector connecting the centre of the nanoparticle and the
centre of mass of the chain. The same is done for the smallest span, $W_1$. Plotted in Figure \ref{fig:spans_orientation}
are the order parameters for the shortest and longest span as functions of the chain centre of mass position.
In the same fashion, $S_L$ is the order parameter of the eigenvectors corresponding to the greater and smaller 
eigenvalues (Figure \ref{fig:eigenvalues_orientation}).
In the interfacial region, we observe a strong tendency for the longest span to orient parallel to the surface
and the shortest span to orient perpendicular to the surface of the nanoparticle. This tendency remains unchanged
if, instead of spans, the principal axes of inertia (i.e., the eigenvectors of the radius of gyration tensor) are
used as measures of shape (Figure \ref{fig:eigenvalues_orientation}). The reorientation of chains is confined
to a region whose width is commensurate with $R_g$.  

\subsection{Chain conformation in the presence of many nanoparticles}

\begin{figure}
\centering
\includegraphics[width=\columnwidth]{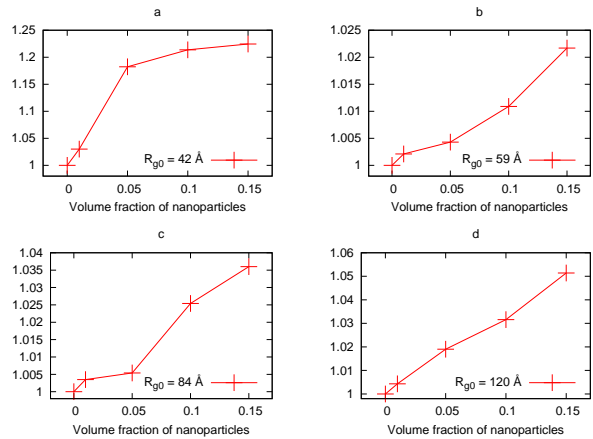}
\caption{Deviation of radius of gyration $R_g$ from its reference value (melt without nanoparticles) $R_{g0}$ for 
the four chain lengths under consideration (a,b,c,d for $32$, $64$, $128$ and $256$ Kuhn segments per
polymeric chain respectively).}
\label{fig:rg_comparison}
\end{figure}

The values of radius of gyration $R_g$ relative to the value for the pure polymer melt $R_{g0}$,
are shown in Figure \ref{fig:rg_comparison} as a function of the nanoparticle volume fraction
for the four different chain lengths used in this work. 
In general, an expansion of polymeric chains with increasing
nanoparticle volume fraction is observed for all chain lengths. 
This expansion is maximal for $32$ Kuhn 
segments per chain, where the radius of gyration $R_{g0} = 42\: \text{\AA}$ is comparable to the radius
of the nanoparticle $R_n = 36$ \AA. It seems that there is a tendency of chains 
to swell when their dimension is equal to
or approaches the dimension of the nanoparticle. This observation is in very good quantitive agreement with
experimental data reported for the same system \cite{Tuteja_PRL_2008}.
In all other cases, the swelling due to the presence of the nanoparticles is hardly distinguished.

\begin{figure}
\centering
\includegraphics[width=\columnwidth]{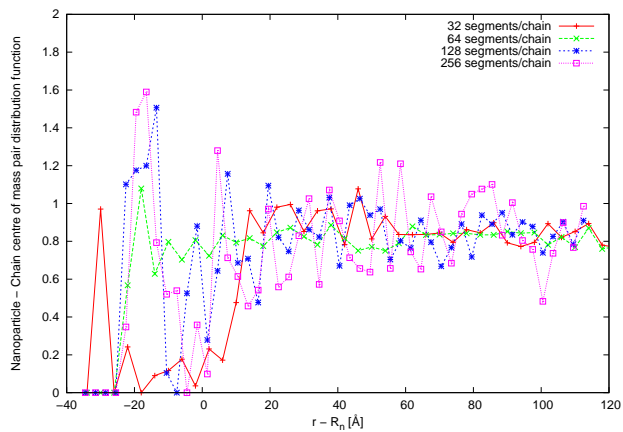}
\caption{Chain centre of mass distribution, accumulated in $4.5$-\AA-thick bins.
The system under consideration consist of dispersed nanoparticles of $36 \:\text{\AA}$ at volume fraction 15 \%. }
\label{fig:cms_many}
\end{figure}

In the presence of many nanoparticles, structural features may be revealed by probing the pair distribution function
between the nanoparticles and the polymer chain centres of mass.  
The COM of the chains are accumulated in $4.5$-\AA-thick bins (Figure \ref{fig:cms_many}).
The pair distribution function depicted, is indicative of the engulfment of nanoparticles by the polymeric chains. 
There are polymeric chains of every chain length, whose COM position lies inside the nanoparticle.
There is a remarkable difference between Figure \ref{fig:cms_many} (many nanoparticles) 
and Figure \ref{fig:cms_around_particle} (one nanoparticle). 
In Figure \ref{fig:cms_around_particle}, the COMs of longer chains are more depleted in the space occupied by the nanoparticle. 
In Figure \ref{fig:cms_many}, the shortest chains of $N=32$ exhibit a tendency to surround the nanoparticles,  
having a clear maximum in their COM distribution inside the nanoparticle, very close to its centre, 
also outside the nanoparticle, some distance from its surface, with a minimum in between.
Interestingly, with  increasing distance the nanoparticle-chain centre of mass distribution goes through a minimum 
and then rises again a short distance away from the nanoparticle surface, because of the contribution of chains which 
are adsorbed to, but do not engulf the nanoparticle.

\section{Conclusions}
\par
The structure of a polymer matrix surrounding an immersed spherical nanoparticle was studied in 
detail both at the segment and the chain
levels. The simulation method employed relies on a hybrid Monte Carlo
formalism. It is a field theoretic approach, based on the grounds of
the self-consistent field, which can treat efficiently large systems.
The system investigated bears great resemblance with an athermal,
all-polystyrene nanocomposite material that has been studied experimentally.
Several conformational properties of the system have been examined. At
the segment level, an increase of the local polymer density
around the nanoparticle is found. Density profiles are weakly dependent on the chain
length. An increase of the degree of polymerization of the chain
results in slightly increased thickness of the depletion layer. The bonds tend
to orient parallel to the nanoparticle surface, as revealed by the second
Legendre polynomial. 
All these effects have a characteristic length scale commensurate with the chain segment size.
\par 
At the chain level, chain centres of mass are able to penetrate the interior of the particle, 
as chains engulf the particle.
This phenomenon is more pronounced when the radius of gyration of chains is comparable in size to the nanoparticle.  
The formation of chain layers around the particle is
suggested by the particle-polymer centre of mass pair distribution
function. The size and the shape of the chains are also affected. A flattening of 
polymer molecules when their centre of mass is close to the nanoparticle centre is revealed by both the chain 
spans and the eigenvalues of the radius of gyration tensor. This
deformation of the polymer  molecules is small for long chains.
\par
In the case of the dispersion of many nanoparticles, deformation of the chain shape is observed.
The chains which share the same length scale with the dispersed particles tend to swell.
This is in agreement with recent neutron scattering measurements.
The radius of gyration increases with increased nanoparticle volume fraction.

\section*{Acknowledgments}
This work was funded by the European Union under the FP7-NMP-2007
program, Nanomodel Grant Agreement SL-208-211778. 
Computational work was carried out under the HPC-EUROPA2 project (project number: 228398) 
with the support of the European Commission Capacities Area - Research Infrastructures Initiative.
Fruitful discussions with Dr. Kostas Daoulas are gratefully acknowledged.




\section*{References}
\bibliography{coarse}

\providecommand{\latin}[1]{#1}
\providecommand*\mcitethebibliography{\thebibliography}
\csname @ifundefined\endcsname{endmcitethebibliography}
  {\let\endmcitethebibliography\endthebibliography}{}
\begin{mcitethebibliography}{49}
\providecommand*\natexlab[1]{#1}
\providecommand*\mciteSetBstSublistMode[1]{}
\providecommand*\mciteSetBstMaxWidthForm[2]{}
\providecommand*\mciteBstWouldAddEndPuncttrue
  {\def\EndOfBibitem{\unskip.}}
\providecommand*\mciteBstWouldAddEndPunctfalse
  {\let\EndOfBibitem\relax}
\providecommand*\mciteSetBstMidEndSepPunct[3]{}
\providecommand*\mciteSetBstSublistLabelBeginEnd[3]{}
\providecommand*\EndOfBibitem{}
\mciteSetBstSublistMode{f}
\mciteSetBstMaxWidthForm{subitem}{(\alph{mcitesubitemcount})}
\mciteSetBstSublistLabelBeginEnd
  {\mcitemaxwidthsubitemform\space}
  {\relax}
  {\relax}

\bibitem[Gersappe(2002)]{Gersappe_PRL_2002}
Gersappe,~D. \emph{Phys. Rev. Lett.} \textbf{2002}, \emph{89}, 058301\relax
\mciteBstWouldAddEndPuncttrue
\mciteSetBstMidEndSepPunct{\mcitedefaultmidpunct}
{\mcitedefaultendpunct}{\mcitedefaultseppunct}\relax
\EndOfBibitem
\bibitem[Balazs \latin{et~al.}(2006)Balazs, Emrick, and
  Russel]{Balazs_Science_2006}
Balazs,~A.~C.; Emrick,~T.; Russel,~T.~P. \emph{Science} \textbf{2006},
  \emph{314}, 1107\relax
\mciteBstWouldAddEndPuncttrue
\mciteSetBstMidEndSepPunct{\mcitedefaultmidpunct}
{\mcitedefaultendpunct}{\mcitedefaultseppunct}\relax
\EndOfBibitem
\bibitem[Bansal \latin{et~al.}(2006)Bansal, Yang, Li, Benicewiz, Kumar, and
  Schadler]{Kumar_2006}
Bansal,~A.; Yang,~H.; Li,~C.; Benicewiz,~B.~C.; Kumar,~S.~K.; Schadler,~L.~S.
  \emph{J. Polym. Sci., Part B: Polym. Phys.} \textbf{2006}, \emph{44},
  2944--2950\relax
\mciteBstWouldAddEndPuncttrue
\mciteSetBstMidEndSepPunct{\mcitedefaultmidpunct}
{\mcitedefaultendpunct}{\mcitedefaultseppunct}\relax
\EndOfBibitem
\bibitem[Lee \latin{et~al.}(2004)Lee, Buxton, and Balazs]{Youn_JChemPhys_2004}
Lee,~J.~Y.; Buxton,~G.~A.; Balazs,~A.~C. \emph{J. Chem. Phys.} \textbf{2004},
  \emph{121}, 5531--5540\relax
\mciteBstWouldAddEndPuncttrue
\mciteSetBstMidEndSepPunct{\mcitedefaultmidpunct}
{\mcitedefaultendpunct}{\mcitedefaultseppunct}\relax
\EndOfBibitem
\bibitem[Bockstaller and Thomas(2004)Bockstaller, and
  Thomas]{Bockstaller_PRL_2004}
Bockstaller,~M.~R.; Thomas,~E.~L. \emph{Phys. Rev. Lett.} \textbf{2004},
  \emph{93}, 166106\relax
\mciteBstWouldAddEndPuncttrue
\mciteSetBstMidEndSepPunct{\mcitedefaultmidpunct}
{\mcitedefaultendpunct}{\mcitedefaultseppunct}\relax
\EndOfBibitem
\bibitem[Si \latin{et~al.}(2006)Si, Araki, Ade, Kilcoyne, Fisher, Sokolov, and
  Rafailovich]{Si_Macromolecules_2006}
Si,~M.; Araki,~T.; Ade,~H.; Kilcoyne,~A. L.~D.; Fisher,~R.; Sokolov,~J.~C.;
  Rafailovich,~M.~H. \emph{Macromolecules} \textbf{2006}, \emph{39}\relax
\mciteBstWouldAddEndPuncttrue
\mciteSetBstMidEndSepPunct{\mcitedefaultmidpunct}
{\mcitedefaultendpunct}{\mcitedefaultseppunct}\relax
\EndOfBibitem
\bibitem[Stratford \latin{et~al.}(2005)Stratford, Adhikari, Pagonabarraga,
  Desplat, and Cates]{Stratford_Science_2005}
Stratford,~K.; Adhikari,~R.; Pagonabarraga,~I.; Desplat,~J.-C.; Cates,~M.~E.
  \emph{Science} \textbf{2005}, \emph{309}, 2198--2201\relax
\mciteBstWouldAddEndPuncttrue
\mciteSetBstMidEndSepPunct{\mcitedefaultmidpunct}
{\mcitedefaultendpunct}{\mcitedefaultseppunct}\relax
\EndOfBibitem
\bibitem[Tuteja \latin{et~al.}(2008)Tuteja, Duxbury, and
  Mackay]{Tuteja_PRL_2008}
Tuteja,~A.; Duxbury,~P.~M.; Mackay,~M.~E. \emph{Phys. Rev. Lett.}
  \textbf{2008}, \emph{100}, 077801\relax
\mciteBstWouldAddEndPuncttrue
\mciteSetBstMidEndSepPunct{\mcitedefaultmidpunct}
{\mcitedefaultendpunct}{\mcitedefaultseppunct}\relax
\EndOfBibitem
\bibitem[Mackay \latin{et~al.}(2006)Mackay, Tuteja, Duxbury, Hawker, Horn,
  Guan, Chen, and Krishman]{Mackay_Science_2006}
Mackay,~M.~E.; Tuteja,~A.; Duxbury,~P.~M.; Hawker,~C.~J.; Horn,~B.~V.;
  Guan,~Z.; Chen,~G.; Krishman,~R. \emph{Science} \textbf{2006}, \emph{311},
  1740\relax
\mciteBstWouldAddEndPuncttrue
\mciteSetBstMidEndSepPunct{\mcitedefaultmidpunct}
{\mcitedefaultendpunct}{\mcitedefaultseppunct}\relax
\EndOfBibitem
\bibitem[Mackay \latin{et~al.}(2003)Mackay, Dao, Tuteja, Ho, Van~Horn, and
  Kim]{Mackay_NatureMaterials_2003}
Mackay,~M.~E.; Dao,~T.~T.; Tuteja,~A.; Ho,~D.~L.; Van~Horn,~B.;
  Kim,~C.~J.,~Ho-Cheol amd~Hawker \emph{Nat. Mater.} \textbf{2003}, \emph{2},
  762--766\relax
\mciteBstWouldAddEndPuncttrue
\mciteSetBstMidEndSepPunct{\mcitedefaultmidpunct}
{\mcitedefaultendpunct}{\mcitedefaultseppunct}\relax
\EndOfBibitem
\bibitem[Doxastakis \latin{et~al.}(2004)Doxastakis, Chen, Guzman, and
  de~Pablo]{Doxastakis_JChemPhys_2004}
Doxastakis,~M.; Chen,~Y.-L.; Guzman,~O.; de~Pablo,~J.~J. \emph{J. Chem. Phys.}
  \textbf{2004}, \emph{120}, 9335--9342\relax
\mciteBstWouldAddEndPuncttrue
\mciteSetBstMidEndSepPunct{\mcitedefaultmidpunct}
{\mcitedefaultendpunct}{\mcitedefaultseppunct}\relax
\EndOfBibitem
\bibitem[Doxastakis \latin{et~al.}(2005)Doxastakis, Chen, and
  de~Pablo]{Doxastakis_JChemPhys_2005}
Doxastakis,~M.; Chen,~Y.-L.; de~Pablo,~J.~J. \emph{J. Chem. Phys.}
  \textbf{2005}, \emph{123}, 034901\relax
\mciteBstWouldAddEndPuncttrue
\mciteSetBstMidEndSepPunct{\mcitedefaultmidpunct}
{\mcitedefaultendpunct}{\mcitedefaultseppunct}\relax
\EndOfBibitem
\bibitem[Hall \latin{et~al.}(2010)Hall, Jayaraman, and
  Schweizer]{Hall_CurrentOpinion_2010}
Hall,~L.~M.; Jayaraman,~A.; Schweizer,~K.~S. \emph{Curr. Opin. Solid State
  Mater. Sci.} \textbf{2010}, \emph{14}, 38 -- 48, Polymers\relax
\mciteBstWouldAddEndPuncttrue
\mciteSetBstMidEndSepPunct{\mcitedefaultmidpunct}
{\mcitedefaultendpunct}{\mcitedefaultseppunct}\relax
\EndOfBibitem
\bibitem[de~Luzuriaga \latin{et~al.}(2009)de~Luzuriaga, Grande, and
  Pomposo]{Luzuriaga_JChemPhys_2009}
de~Luzuriaga,~A.~R.; Grande,~H.~J.; Pomposo,~J.~A. \emph{J. Chem. Phys.}
  \textbf{2009}, \emph{130}, 084905\relax
\mciteBstWouldAddEndPuncttrue
\mciteSetBstMidEndSepPunct{\mcitedefaultmidpunct}
{\mcitedefaultendpunct}{\mcitedefaultseppunct}\relax
\EndOfBibitem
\bibitem[Pomposo \latin{et~al.}(2008)Pomposo, de~Luzuriaga, Etxeberria, and
  Rodríguez]{Pomposo_PCCP_2008}
Pomposo,~J.~A.; de~Luzuriaga,~A.~R.; Etxeberria,~A.; Rodríguez,~J. \emph{Phys.
  Chem. Chem. Phys.} \textbf{2008}, \emph{10}, 650\relax
\mciteBstWouldAddEndPuncttrue
\mciteSetBstMidEndSepPunct{\mcitedefaultmidpunct}
{\mcitedefaultendpunct}{\mcitedefaultseppunct}\relax
\EndOfBibitem
\bibitem[Arthi~Jayaraman(2008)]{Jayaraman_Macromolecules_2008}
Arthi~Jayaraman,~K. S.~S. \emph{Macromolecules} \textbf{2008}, \emph{41},
  9430--9438\relax
\mciteBstWouldAddEndPuncttrue
\mciteSetBstMidEndSepPunct{\mcitedefaultmidpunct}
{\mcitedefaultendpunct}{\mcitedefaultseppunct}\relax
\EndOfBibitem
\bibitem[Hall and Schweizer(2008)Hall, and Schweizer]{Hall_JChemPhys_2008}
Hall,~L.~M.; Schweizer,~K.~S. \emph{J. Chem. Phys.} \textbf{2008}, \emph{128},
  234901\relax
\mciteBstWouldAddEndPuncttrue
\mciteSetBstMidEndSepPunct{\mcitedefaultmidpunct}
{\mcitedefaultendpunct}{\mcitedefaultseppunct}\relax
\EndOfBibitem
\bibitem[Fredrickson(2006)]{Fredrickson_EquilibriumTheory}
Fredrickson,~G.~H. \emph{The equilibrium theory of inhomogeneous polymers};
  Clarendon Press, 2006\relax
\mciteBstWouldAddEndPuncttrue
\mciteSetBstMidEndSepPunct{\mcitedefaultmidpunct}
{\mcitedefaultendpunct}{\mcitedefaultseppunct}\relax
\EndOfBibitem
\bibitem[Surve \latin{et~al.}(2006)Surve, Pryamitsyn, and
  Ganesan]{Surve_Langmuir_2006}
Surve,~M.; Pryamitsyn,~V.; Ganesan,~V. \emph{Langmuir} \textbf{2006},
  \emph{22}, 969--981\relax
\mciteBstWouldAddEndPuncttrue
\mciteSetBstMidEndSepPunct{\mcitedefaultmidpunct}
{\mcitedefaultendpunct}{\mcitedefaultseppunct}\relax
\EndOfBibitem
\bibitem[Surve \latin{et~al.}(2007)Surve, Pryamitsyn, and
  Ganesan]{Surve_Macromolecules_2007}
Surve,~M.; Pryamitsyn,~V.; Ganesan,~V. \emph{Macromolecules} \textbf{2007},
  \emph{40}, 344--354\relax
\mciteBstWouldAddEndPuncttrue
\mciteSetBstMidEndSepPunct{\mcitedefaultmidpunct}
{\mcitedefaultendpunct}{\mcitedefaultseppunct}\relax
\EndOfBibitem
\bibitem[Ganesan \latin{et~al.}(2008)Ganesan, Khounlavong, and
  Pryamitsyn]{Ganesan_PhysRevE_2008}
Ganesan,~V.; Khounlavong,~L.; Pryamitsyn,~V. \emph{Phys. Rev. E} \textbf{2008},
  \emph{78}, 051804\relax
\mciteBstWouldAddEndPuncttrue
\mciteSetBstMidEndSepPunct{\mcitedefaultmidpunct}
{\mcitedefaultendpunct}{\mcitedefaultseppunct}\relax
\EndOfBibitem
\bibitem[Harton and Kumar(2008)Harton, and Kumar]{Harton_PolymerScience_2008}
Harton,~S.~E.; Kumar,~S.~K. \emph{J. Polym. Sci., Part B: Polym. Phys.}
  \textbf{2008}, \emph{46}, 351--358\relax
\mciteBstWouldAddEndPuncttrue
\mciteSetBstMidEndSepPunct{\mcitedefaultmidpunct}
{\mcitedefaultendpunct}{\mcitedefaultseppunct}\relax
\EndOfBibitem
\bibitem[Sides \latin{et~al.}(2006)Sides, Kim, Kramer, and
  Fredrickson]{Sides_PRL_2006}
Sides,~S.~W.; Kim,~B.~J.; Kramer,~E.~J.; Fredrickson,~G.~H. \emph{Phys. Rev.
  Lett.} \textbf{2006}, \emph{96}, 250601\relax
\mciteBstWouldAddEndPuncttrue
\mciteSetBstMidEndSepPunct{\mcitedefaultmidpunct}
{\mcitedefaultendpunct}{\mcitedefaultseppunct}\relax
\EndOfBibitem
\bibitem[Thompson \latin{et~al.}(2002)Thompson, Ginzburg, Matsen, and
  Balazs]{Thompson_Macromolecules_2002}
Thompson,~R.~B.; Ginzburg,~V.~V.; Matsen,~M.~W.; Balazs,~A.~C.
  \emph{Macromolecules} \textbf{2002}, \emph{35}, 1060--1071\relax
\mciteBstWouldAddEndPuncttrue
\mciteSetBstMidEndSepPunct{\mcitedefaultmidpunct}
{\mcitedefaultendpunct}{\mcitedefaultseppunct}\relax
\EndOfBibitem
\bibitem[Buxton \latin{et~al.}(2003)Buxton, Lee, and
  Balazs]{Buxton_Macromolecules_2003}
Buxton,~G.~A.; Lee,~J.~Y.; Balazs,~A.~C. \emph{Macromolecules} \textbf{2003},
  \emph{36}, 9631--9637\relax
\mciteBstWouldAddEndPuncttrue
\mciteSetBstMidEndSepPunct{\mcitedefaultmidpunct}
{\mcitedefaultendpunct}{\mcitedefaultseppunct}\relax
\EndOfBibitem
\bibitem[Lee \latin{et~al.}(2003)Lee, Shou, and Balazs]{Lee_PRL_2003}
Lee,~J.~Y.; Shou,~Z.; Balazs,~A.~C. \emph{Phys. Rev. Lett.} \textbf{2003},
  \emph{91}, 136103\relax
\mciteBstWouldAddEndPuncttrue
\mciteSetBstMidEndSepPunct{\mcitedefaultmidpunct}
{\mcitedefaultendpunct}{\mcitedefaultseppunct}\relax
\EndOfBibitem
\bibitem[Daoulas and M{\"u}ller(2006)Daoulas, and
  M{\"u}ller]{DaoulasJChemPhys_2006}
Daoulas,~K.~C.; M{\"u}ller,~M. \emph{J.Chem.Phys.} \textbf{2006}, \emph{125},
  184904\relax
\mciteBstWouldAddEndPuncttrue
\mciteSetBstMidEndSepPunct{\mcitedefaultmidpunct}
{\mcitedefaultendpunct}{\mcitedefaultseppunct}\relax
\EndOfBibitem
\bibitem[Daoulas \latin{et~al.}(2006)Daoulas, M{\"u}ller, Pablo, Nealey, and
  Smith]{DaoulasSoftMatter_2006}
Daoulas,~K.~C.; M{\"u}ller,~M.; Pablo,~J.; Nealey,~P.; Smith,~G. \emph{Soft
  Matter} \textbf{2006}, \emph{2}, 573--583\relax
\mciteBstWouldAddEndPuncttrue
\mciteSetBstMidEndSepPunct{\mcitedefaultmidpunct}
{\mcitedefaultendpunct}{\mcitedefaultseppunct}\relax
\EndOfBibitem
\bibitem[Detcheverry \latin{et~al.}(2009)Detcheverry, Pike, Nealey, M\"uller,
  and de~Pablo]{Detcheverry_PRL_2009}
Detcheverry,~F.~A.; Pike,~D.~Q.; Nealey,~P.~F.; M\"uller,~M.; de~Pablo,~J.~J.
  \emph{Phys. Rev. Lett.} \textbf{2009}, \emph{102}, 197801\relax
\mciteBstWouldAddEndPuncttrue
\mciteSetBstMidEndSepPunct{\mcitedefaultmidpunct}
{\mcitedefaultendpunct}{\mcitedefaultseppunct}\relax
\EndOfBibitem
\bibitem[Detcheverry \latin{et~al.}(2008)Detcheverry, Kang, Daoulas, M\"uller,
  Nealey, and de~Pablo]{Daoulas_Macromolecules_2008}
Detcheverry,~F.~A.; Kang,~H.; Daoulas,~K.~C.; M\"uller,~M.; Nealey,~P.~F.;
  de~Pablo,~J.~J. \emph{Macromolecules} \textbf{2008}, \emph{41},
  4989--5001\relax
\mciteBstWouldAddEndPuncttrue
\mciteSetBstMidEndSepPunct{\mcitedefaultmidpunct}
{\mcitedefaultendpunct}{\mcitedefaultseppunct}\relax
\EndOfBibitem
\bibitem[Stoykovich \latin{et~al.}(2010)Stoykovich, Daoulas, M{\"u}ller, Kang,
  de~Pablo, and Nealey]{Stoykovich_Macromolecules_2010}
Stoykovich,~M.~P.; Daoulas,~K.~C.; M{\"u}ller,~M.; Kang,~H.; de~Pablo,~J.~J.;
  Nealey,~P.~F. \emph{Macromolecules} \textbf{2010}, \emph{43},
  2334--2342\relax
\mciteBstWouldAddEndPuncttrue
\mciteSetBstMidEndSepPunct{\mcitedefaultmidpunct}
{\mcitedefaultendpunct}{\mcitedefaultseppunct}\relax
\EndOfBibitem
\bibitem[Helfand and Tagami(1972)Helfand, and
  Tagami]{Helfand_Tagami_JChemPhys_1972a}
Helfand,~E.; Tagami,~Y. \emph{J.Chem.Phys.} \textbf{1972}, \emph{56},
  3592\relax
\mciteBstWouldAddEndPuncttrue
\mciteSetBstMidEndSepPunct{\mcitedefaultmidpunct}
{\mcitedefaultendpunct}{\mcitedefaultseppunct}\relax
\EndOfBibitem
\bibitem[Dodd and Theodorou(1991)Dodd, and
  Theodorou]{DoddTheodorou_MolPhys_1991}
Dodd,~L.~R.; Theodorou,~D.~N. \emph{Mol.Phys.} \textbf{1991}, \emph{72},
  1313\relax
\mciteBstWouldAddEndPuncttrue
\mciteSetBstMidEndSepPunct{\mcitedefaultmidpunct}
{\mcitedefaultendpunct}{\mcitedefaultseppunct}\relax
\EndOfBibitem
\bibitem[Laradji \latin{et~al.}(1994)Laradji, Guo, and
  Zuckermann]{Laradji_PhysRevE_1994}
Laradji,~M.; Guo,~H.; Zuckermann,~M.~J. \emph{Phys. Rev. E} \textbf{1994},
  \emph{49}, 3199--3206\relax
\mciteBstWouldAddEndPuncttrue
\mciteSetBstMidEndSepPunct{\mcitedefaultmidpunct}
{\mcitedefaultendpunct}{\mcitedefaultseppunct}\relax
\EndOfBibitem
\bibitem[Hiemenz(1977)]{Hiemenz}
Hiemenz,~P.~C. \emph{Principles of colloid and surface chemistry}, 2nd ed.;
  Dekker, 1977\relax
\mciteBstWouldAddEndPuncttrue
\mciteSetBstMidEndSepPunct{\mcitedefaultmidpunct}
{\mcitedefaultendpunct}{\mcitedefaultseppunct}\relax
\EndOfBibitem
\bibitem[Hamaker(1937)]{Hamaker_original}
Hamaker,~H. \emph{Physica(Amsterdam)} \textbf{1937}, \emph{IV}, 1058\relax
\mciteBstWouldAddEndPuncttrue
\mciteSetBstMidEndSepPunct{\mcitedefaultmidpunct}
{\mcitedefaultendpunct}{\mcitedefaultseppunct}\relax
\EndOfBibitem
\bibitem[Everaers and Ejtehadi(2003)Everaers, and Ejtehadi]{Everaers_2003}
Everaers,~R.; Ejtehadi,~M. \emph{Phys.Rev.E} \textbf{2003}, \emph{67},
  041710\relax
\mciteBstWouldAddEndPuncttrue
\mciteSetBstMidEndSepPunct{\mcitedefaultmidpunct}
{\mcitedefaultendpunct}{\mcitedefaultseppunct}\relax
\EndOfBibitem
\bibitem[Spyriouni \latin{et~al.}(2007)Spyriouni, Tzoumanekas, Theodorou,
  Müller-Plathe, and Milano]{Spyriouni_Macromolecules_2007}
Spyriouni,~T.; Tzoumanekas,~C.; Theodorou,~D.; Müller-Plathe,~F.; Milano,~G.
  \emph{Macromolecules} \textbf{2007}, \emph{40}, 3876--3885\relax
\mciteBstWouldAddEndPuncttrue
\mciteSetBstMidEndSepPunct{\mcitedefaultmidpunct}
{\mcitedefaultendpunct}{\mcitedefaultseppunct}\relax
\EndOfBibitem
\bibitem[Good and Hope(1971)Good, and Hope]{GoodHope_1971}
Good,~R.~J.; Hope,~C.~J. \emph{J.Chem.Phys.} \textbf{1971}, \emph{55},
  111\relax
\mciteBstWouldAddEndPuncttrue
\mciteSetBstMidEndSepPunct{\mcitedefaultmidpunct}
{\mcitedefaultendpunct}{\mcitedefaultseppunct}\relax
\EndOfBibitem
\bibitem[Marsaglia(1972)]{Marsaglia_AnnMathStat_1972}
Marsaglia,~G. \emph{Ann. Math. Stat.} \textbf{1972}, \emph{43}, 645--646\relax
\mciteBstWouldAddEndPuncttrue
\mciteSetBstMidEndSepPunct{\mcitedefaultmidpunct}
{\mcitedefaultendpunct}{\mcitedefaultseppunct}\relax
\EndOfBibitem
\bibitem[Auhl \latin{et~al.}(2003)Auhl, Everaers, Grest, Kremer, and
  Plimpton]{AuhlEveraers_JChemPhys_2003}
Auhl,~R.; Everaers,~R.; Grest,~G.~S.; Kremer,~K.; Plimpton,~S.~J. \emph{J.
  Chem. Phys.} \textbf{2003}, \emph{119}, 12718\relax
\mciteBstWouldAddEndPuncttrue
\mciteSetBstMidEndSepPunct{\mcitedefaultmidpunct}
{\mcitedefaultendpunct}{\mcitedefaultseppunct}\relax
\EndOfBibitem
\bibitem[Aarts \latin{et~al.}(2002)Aarts, Tuinier, and
  Lekkerkerker]{Aarts_JPhysCondMatter_2002}
Aarts,~D. G.~L.; Tuinier,~R.; Lekkerkerker,~H. N.~W. \emph{J. Phys.: Condens.
  Matter} \textbf{2002}, \emph{14}, 7551--7561\relax
\mciteBstWouldAddEndPuncttrue
\mciteSetBstMidEndSepPunct{\mcitedefaultmidpunct}
{\mcitedefaultendpunct}{\mcitedefaultseppunct}\relax
\EndOfBibitem
\bibitem[Brown \latin{et~al.}(2008)Brown, Marcadon, Mélé, and
  Albérola]{Brown_Macromolecules_2008}
Brown,~D.; Marcadon,~V.; Mélé,~P.; Albérola,~N.~D. \emph{Macromolecules}
  \textbf{2008}, \emph{41}, 1499--1511\relax
\mciteBstWouldAddEndPuncttrue
\mciteSetBstMidEndSepPunct{\mcitedefaultmidpunct}
{\mcitedefaultendpunct}{\mcitedefaultseppunct}\relax
\EndOfBibitem
\bibitem[Rubin and J.Mazur(1977)Rubin, and J.Mazur]{RubinMazur_1977}
Rubin,~J.; J.Mazur, \emph{Macromolecules} \textbf{1977}, \emph{10},
  139--149\relax
\mciteBstWouldAddEndPuncttrue
\mciteSetBstMidEndSepPunct{\mcitedefaultmidpunct}
{\mcitedefaultendpunct}{\mcitedefaultseppunct}\relax
\EndOfBibitem
\bibitem[Theodorou and Suter(1985)Theodorou, and Suter]{TheodorouSuter_1985}
Theodorou,~D.~N.; Suter,~U.~W. \emph{Macromolecules} \textbf{1985}, \emph{18},
  1206--1214\relax
\mciteBstWouldAddEndPuncttrue
\mciteSetBstMidEndSepPunct{\mcitedefaultmidpunct}
{\mcitedefaultendpunct}{\mcitedefaultseppunct}\relax
\EndOfBibitem
\bibitem[Picu and Ozmusul(2003)Picu, and Ozmusul]{PicuOzmusul_2003}
Picu,~R.~C.; Ozmusul,~M.~S. \emph{J. Chem. Phys.} \textbf{2003}, \emph{118},
  11239--11248\relax
\mciteBstWouldAddEndPuncttrue
\mciteSetBstMidEndSepPunct{\mcitedefaultmidpunct}
{\mcitedefaultendpunct}{\mcitedefaultseppunct}\relax
\EndOfBibitem
\bibitem[Mansfield and Theodorou(1990)Mansfield, and
  Theodorou]{TheodorouMansfield_1990}
Mansfield,~K.~F.; Theodorou,~D.~N. \emph{Macromolecules} \textbf{1990},
  \emph{23}, 4430--4445\relax
\mciteBstWouldAddEndPuncttrue
\mciteSetBstMidEndSepPunct{\mcitedefaultmidpunct}
{\mcitedefaultendpunct}{\mcitedefaultseppunct}\relax
\EndOfBibitem
\bibitem[Mansfield and Theodorou(1991)Mansfield, and
  Theodorou]{TheodorouMansfield_1991}
Mansfield,~K.~F.; Theodorou,~D.~N. \emph{Macromolecules} \textbf{1991},
  \emph{24}, 4295--4309\relax
\mciteBstWouldAddEndPuncttrue
\mciteSetBstMidEndSepPunct{\mcitedefaultmidpunct}
{\mcitedefaultendpunct}{\mcitedefaultseppunct}\relax
\EndOfBibitem
\end{mcitethebibliography}







\end{document}